\documentclass[twocolumn]{aastex631}

\newcommand*{\Chandra}{Chandra} 
\defcitealias{Mantz2111.09343}{M22}
\newcommand*{\Mtwentytwo}{\citetalias{Mantz2111.09343}}
\defcitealias{Mantz1509.01322}{M16}
\newcommand*{\Msixteen}{\citetalias{Mantz1509.01322}}
\defcitealias{Darragh-Ford2302.10931}{D-F23}
\newcommand*{\DFtwentythree}{\citetalias{Darragh-Ford2302.10931}}
\usepackage{amsmath}

\shorttitle{Deep Chandra Observations of the $z=1.16$ Relaxed, Cool-Core Galaxy Cluster SPT-CL J2215-3537}
\shortauthors{Stueber et al.}

\graphicspath{{./}{figures/}}

\begin{document}

\title {Deep Chandra Observations of the $z=1.16$ Relaxed, Cool-core Galaxy Cluster SPT-CL J2215-3537}

\author[0000-0002-2776-978X]{Haley R. Stueber}
\affiliation{Department of Physics, Stanford University, 382 Via Pueblo Mall, Stanford, CA 94305, USA}
\affiliation{Kavli Institute for Particle Astrophysics and Cosmology, Stanford University, 452 Lomita Mall, Stanford, CA 94305, USA}

\author[0000-0002-8031-1217]{Adam B. Mantz}
\affiliation{Kavli Institute for Particle Astrophysics and Cosmology, Stanford University, 452 Lomita Mall, Stanford, CA 94305, USA}

\author[0000-0003-0667-5941]{Steven W. Allen}
\affiliation{Department of Physics, Stanford University, 382 Via Pueblo Mall, Stanford, CA 94305, USA}
\affiliation{Kavli Institute for Particle Astrophysics and Cosmology, Stanford University, 452 Lomita Mall, Stanford, CA 94305, USA}
\affiliation{SLAC National Accelerator Laboratory, 2575 Sand Hill Road, Menlo Park, CA  94025, USA}

\author[0000-0001-7179-6198]{Anthony M. Flores}
\affiliation{Department of Physics, Stanford University, 382 Via Pueblo Mall, Stanford, CA 94305, USA}
\affiliation{Kavli Institute for Particle Astrophysics and Cosmology, Stanford University, 452 Lomita Mall, Stanford, CA 94305, USA}
\affiliation{Department of Physics and Astronomy, Rutgers University, 136 Frelinghuysen Road, Piscataway, NJ 08854, USA}

\author[0000-0003-2985-9962]{R. Glenn Morris}
\affiliation{Department of Physics, Stanford University, 382 Via Pueblo Mall, Stanford, CA 94305, USA}
\affiliation{Kavli Institute for Particle Astrophysics and Cosmology, Stanford University, 452 Lomita Mall, Stanford, CA 94305, USA}
\affiliation{SLAC National Accelerator Laboratory, 2575 Sand Hill Road, Menlo Park, CA  94025, USA}

\author[0009-0001-9176-8861]{Abigail Y. Pan}
\affiliation{Department of Physics, Stanford University, 382 Via Pueblo Mall, Stanford, CA 94305, USA}
\affiliation{Kavli Institute for Particle Astrophysics and Cosmology, Stanford University, 452 Lomita Mall, Stanford, CA 94305, USA}

\author[0000-0003-3521-3631]{Taweewat Somboonpanyakul}
\affiliation{Kavli Institute for Particle Astrophysics and Cosmology, Stanford University, 452 Lomita Mall, Stanford, CA 94305, USA}
\affiliation{Department of Physics, Faculty of Science, Chulalongkorn University, 254 Phyathai Road, Patumwan, Bangkok 10330, Thailand}

\author[0000-0001-7665-5079]{Lindsey E. Bleem}
\affiliation{High Energy Physics Division, Argonne National Laboratory, 9700 S. Cass Avenue, Argonne, IL 60439, USA}
\affiliation{Kavli Institute for Cosmological Physics, University of Chicago, 5640 South Ellis Avenue, Chicago, IL 60637, USA}

\author[0000-0002-2238-2105]{Michael Calzadilla}
\affiliation{Center for Astrophysics | Harvard \& Smithsonian, 60 Garden Street, Cambridge, MA 02138, USA}

\author[0000-0003-4175-571X]{Benjamin Floyd}
\affiliation{Insititute of Cosmology and Gravitation, University of Portsmouth, Dennis Sciama Building, Portsmouth, PO1 3FX, UK}

\author[0000-0001-7271-7340]{Julie Hlavacek-Larrondo}
\affiliation{D\'{e}partement de Physique, Universit\'{e} de Montr\'{e}al, Succ. Centre-Ville, 
Montr\'{e}al, Qu\'{e}bec, H3C 3J7, Canada}

\author[0000-0001-5226-8349]{Michael McDonald}
\affiliation{Department of Physics, Massachusetts Institute of Technology, 77 Massachusetts Avenue, Cambridge, MA 02139, USA}
\affiliation{Kavli Institute for Astrophysics and Space Research, Massachusetts Institute of Technology, 70 Vassar Street, Cambridge, MA 02139, USA}

\author[0000-0002-5222-1337]{Arnab Sarkar}
\affiliation{Kavli Institute for Astrophysics and Space Research, Massachusetts Institute of Technology, 70 Vassar Street, Cambridge, MA 02139, USA}
\affiliation{Department of Physics, University of Arkansas, 825 W Dickson Street, Fayetteville, AR 72701, USA}

\email{hstueber@stanford.edu}


\begin{abstract}
\noindent
 Galaxy clusters serve as a unique and valuable laboratory for probing cosmological models and understanding astrophysics at the high-mass limit of structure formation. Clusters that are dynamically relaxed are especially useful targets of study because of their morphological and dynamical simplicity. However, at redshifts $z>1$, very few such clusters have been identified. We present results from new \Chandra{} observations of the cluster SPT-CL J2215-3537 (hereafter SPT\,J2215), at $z = 1.16$, the second most distant relaxed, cool-core cluster identified to date. We place constraints on the cluster’s total mass profile and investigate its thermodynamic profiles, scaling relations (gas mass, average temperature, and X-ray luminosity), and metal enrichment, resolving the cool core and providing essential context for the massive starburst seen in its central galaxy. We contextualize the thermodynamic and cosmological properties of the cluster within a sample of well-studied, lower-redshift relaxed systems. In this way, SPT\,J2215 serves as a powerful high-redshift benchmark for understanding the formation of cool cores and the evolution of massive clusters of galaxies.
\end{abstract}


\section{Introduction}
\noindent
Observations of the most dynamically relaxed clusters of galaxies provide a unique probe of the growth of large-scale structure and the properties of dark matter and dark energy \citep{Allen0205007, Allen0405340, Allen0706.0033, Allen1103.4829,Ettori0211335, Ettori0904.2740, Schmidt0405374, Borgani0310794, Rapetti0409574, Mantz1402.6212, Mantz2111.09343, Wan2101.09389}. It is in these systems that, due to their morphological simplicity and approximate states of hydrostatic equilibrium, we can reconstruct the 3D, dark matter-inclusive mass profiles most robustly, and study the hot intracluster medium (ICM) with minimal systematic uncertainties introduced from geometric complexity and projection effects \citep{Nagai0609247, Ansarifard1911.07878}. These studies reveal the detailed thermodynamic and chemical composition of the ICM, providing insight into star formation activity, active galactic nucleus (AGN) feedback, and the formation and distribution of heavy elements, throughout the Universe and across cosmic time \citep{2012NJPh...14e5023M, Fabian1204.4114, Werner1310.7948}. For such systems, the concordance cosmological model, Lambda Cold Dark Matter ($\Lambda\text{CDM}$), also provides well-defined predictions for the distribution of mass within these structures (e.g.\ \citealt{Navarro9611107, Navarro0311231, Bullock9908159, Gao0711.0746, Ludlow2016, Brown2020, Klypin1411.4001, Eckert2022, Deimer2023}). The apparent evolution of the ratio of baryonic mass to CDM in clusters is also a strong probe of cosmology, with hydrodynamic simulations of cluster formation indicating that the gas mass fraction at intermediate-to-large cluster radii should evolve minimally with redshift \citep{Eke9708070, Kay0407058, Nagai0609247, Young1007.0887, Borgani0906.4370, Battaglia1209.4082, Planelles1209.5058}.

High resolution X-ray imaging of the hot ICM is a reliable means to assess its dynamics and thus the dynamical state of clusters \citep{Allen1103.4829, Mantz1502.06020}. Galaxy clusters primarily emit X-rays in the form of line emission and bremsstrahlung radiation, the two-body nature of which causes local density fluctuations to appear with high contrast in surface brightness (e.g.\ shock and cold fronts; \citealt{Markevitch0701821}). On larger scales, cluster merger activity drives asymmetries and bulk motions in the ICM that can persist for several sound-crossing times (typically a few Gyr)  \citep{Borgani0906.4370}. Conversely, a smooth, symmetric and centrally peaked surface brightness distribution, with a high central gas density and relatively short cooling time, are signatures of dynamical relaxation. These bright, central regions commonly also exhibit temperature dips, leading to their designation as ``cool cores" \citep{Fabian1994MNRAS.267..779F, White9707269, Peterson0512549, Hudson0911.0409}. A number of studies propose quantitative measurements which attempt either to identify the presence of central surface brightness peaks (e.g.\ \citealt{Vikhlinin0611438, Santos0802.1445, Bohringer0912.4667}) or to measure bulk asymmetry on various scales (e.g.\ \citealt{Mohr1993ApJ...413..492, Buote9502002, Jeltema0501360, Nurgaliev1309.7044, Rasia1211.7040}) as metrics for determining cluster relaxedness. \citet{Mantz1502.06020} proposes a general and automated morphological analysis tool which defines a combination of symmetry, peakiness, and alignment (SPA) criteria. In brief, these criteria evaluate the sharpness of the cluster surface brightness peak, the symmetry of a series of isophotes about a globally defined cluster center, and the alignment of these isophotes with one another.
This SPA selection method was used to identify a sample of 40 massive, dynamically relaxed clusters, with mean X-ray temperatures $kT>5$\,keV measured within 0.15–1 \( r_{500} \), used for thermodynamic studies in \citet{Mantz1509.01322} (hereafter \Msixteen{}), as well as a more recent, expanded sample which includes four additional clusters that were used for cosmological studies (\citealt{Mantz2111.09343, Darragh-Ford2302.10931}, hereafter \Mtwentytwo{} and \DFtwentythree{}, respectively), which we will refer to throughout this paper. 
The highest redshift cluster in the \Mtwentytwo{} sample, SPT-CL J2215-3537 (hereafter SPT\,J2215), is the subject of this work.

SPT\,J2215 was identified through its Sunyaev--Zel'dovich (SZ) effect signal as part of the 2770 square degree South Pole Telescope (SPT) SPTpol Extended Cluster Survey, and its redshift was spectroscopically determined to be $z=1.160$ (SPT-ECS; \citealt{Bleem1910.04121}). Follow-up \Chandra{} observations of SPT\,J2215 were used to assess the dynamical state of the cluster ICM, revealing it to be the highest redshift relaxed, cool-core cluster then known (\Mtwentytwo{}; \citealt{2023ApJ...947...44C}).\footnote{While a relaxed cluster at higher redshift has recently been identified with SPA metrics (Flores et al.\ in prep), SPT\,J2215 remains the highest redshift cluster with deep enough data to be included in the comparison sample of relaxed, cool-cores.}
These same data also enabled initial constraints to be placed on the thermodynamic properties of the cluster, while optical follow-up measurements provided strong evidence for star formation occurring in its cool core \citep{2023ApJ...947...44C}. 

In this paper, we present new, deeper \Chandra{} observations of SPT\,J2215 used to characterize its thermodynamic and cosmological properties, resolving its cool core on small scales and constraining its density, temperature, mass and metallicity profiles. We compare the cosmological properties of SPT\,J2215 to those of the remainder of the SPA-selected, relaxed, cool-core sample studied in \Mtwentytwo{} and \DFtwentythree{}, and its thermodynamic properties to the subsample of 40 clusters in \Msixteen{}. 

The remainder of this paper is structured as follows: Section~\ref{sec:methods} summarizes the observational data used in this work and describes the data reduction and analysis methods employed to model the cluster and background. The results of this analysis are presented and discussed in Section \ref{sec:analysis}, which is divided into four subsections. Section~\ref{sec:thermo_results} discusses the thermodynamic profiles and Section~\ref{sec:scaling_relations} discusses the scaling relations obtained from this analysis in the context of understanding cooling and feedback within the cluster and evaluating its consistency with self-similarity models. Section~\ref{sec:metalicity} presents the radial metallicity profile for the cluster, and discusses the relevance of such results for studies of  metal enrichment in the Universe. Finally, Section~\ref{sec:cosmo_results} presents new constraints on the dark matter halo concentration and gas mass fraction of SPT\,J2215, which, as the highest redshift cluster for which such constraints exist, provides a new benchmark for testing cosmological models. We summarize these results and conclude in Section \ref{sec:conclusion}.
Throughout this analysis, we assume flat $\Lambda \text{CDM}$ cosmology with parameters $H_0 = 70$\,km\,s$^{-1}$\,Mpc$^{-1}$, $\Omega_m = 0.3$ and $\Omega_{\Lambda} = 0.7$. The physical scale is 8.25 kpc\,arcsec$^{-1}$ and the critical density, $\rho_{\rm crit}(z)$, is $3.43\times10^{-29}\,\mathrm{g}\,\mathrm{cm}^{-3}$ at the redshift of the cluster.

\begin{figure*}[ht!]
  \includegraphics[width=0.49\textwidth]{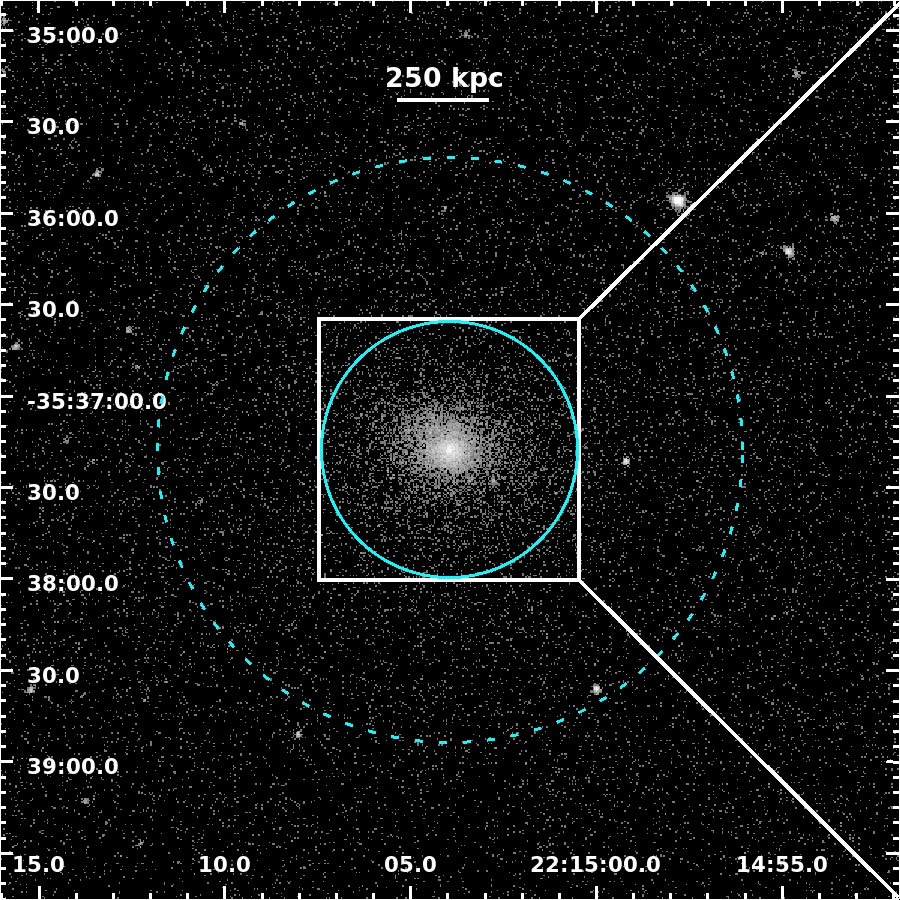}
  \includegraphics[width=0.49\textwidth]{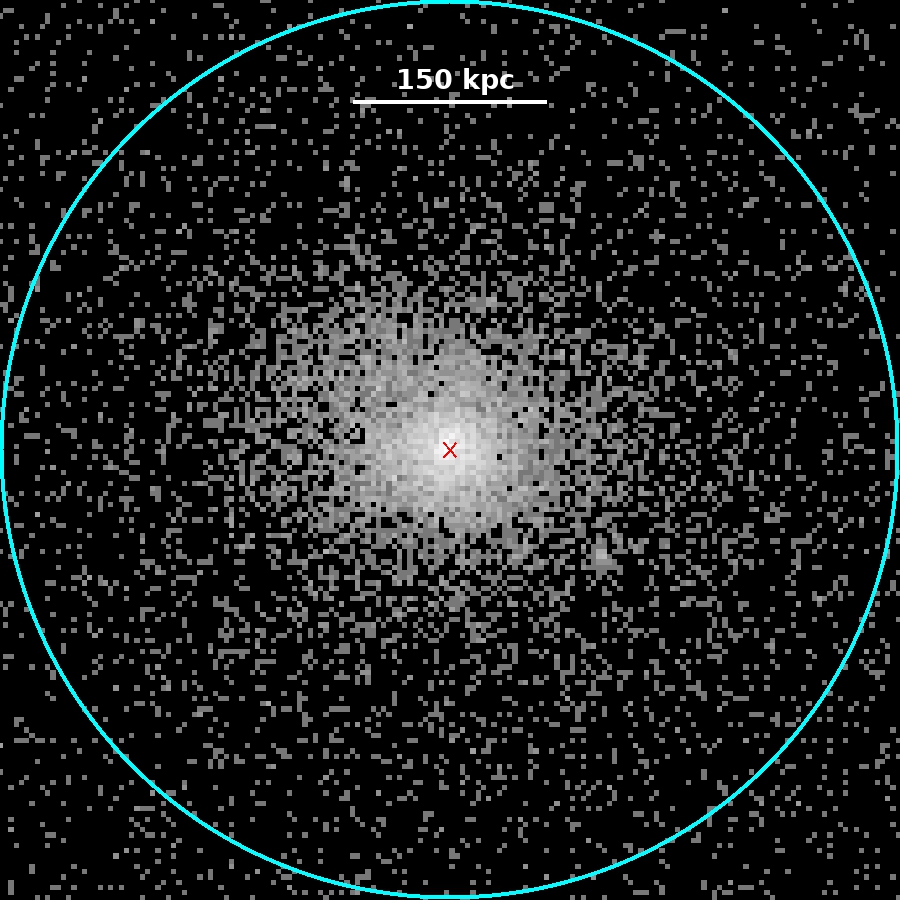}
  \caption{{\it Left:} Broadband (0.6--7.0 keV) Chandra exposure of SPT\,J2215 ($z = 1.16$) totaling $\sim 200$\,ks. The dashed and solid cyan circles show our estimates of $r_{500}$ and $r_{2500}$, respectively. {\it Right:} Zoom-in of the cluster image on the left to the region within $r_{2500}$, where the BCG is indicated by a red `X'.}
  \label{fig:headshots}
\end{figure*}

\section{Data \& Methods}
\label{sec:methods}

\begin{table}[]
    \centering 
    \caption{Summary of Chandra Observations}
    \begin{tabular}{ccccc}
        \hline
        OBSID & Date & Det. & FP Temp. & Clean Time \\
        & & (I/S) & (\(^\circ\)C) & (ks) \\
        \hline
        22653 & 2020-08-17 & I & $-117.93$ & 32.32 \\ 
        24614 & 2020-08-18 & I & $-118.58$ & 26.61 \\ 
        24615 & 2020-08-20 & I & $-119.70$ & 7.93 \\ 
        25468 & 2021-12-24 & S & $-113.73$ & 22.26 \\
        26244 & 2021-12-25 & S & $-113.89$ & 23.24 \\
        25902 & 2022-08-17 & S & $-117.12$ & 28.24 \\
        25903 & 2022-12-19 & S & $-113.57$ & 9.93 \\
        27614 & 2022-12-20 & S & $-112.44$ & 9.93 \\
        25904 & 2023-07-17 & S & $-116.15$ & 30.71 \\
        26372 & 2023-08-28 & S & $-112.27$ & 7.96 \\
    \hline
    \end{tabular}
    \label{tab:obs}
\end{table}

\noindent
An initial set of \Chandra{} ACIS-I observations of SPT\,J2215 were taken in 2020, totaling 66.86\,ks of clean time. We use this data along with 132.27\,ks of new \Chandra{} observations taken with the ACIS-S detector to produce the results presented here. A summary of the observations and their dates, durations, and detectors (``Det.'', either ACIS-I or ACIS-S), along with focal plane temperatures (``FP Temp.'') in degrees Celsius is given in Table \ref{tab:obs}. All observations were made in VFAINT mode. The observations can be found at the \Chandra{} data collection (CDC) 377~\dataset[doi:10.25574/cdc.377]{https://doi.org/10.25574/cdc.377}. Figure~\ref{fig:headshots} shows the broadband, native resolution \Chandra{} image of SPT\,J2215 using all 199.13\,ks of clean observational data. Cyan dashed and solid circles of radius $r_{500}$ and $r_{2500}$ (estimated from this study) designate the radii within which the mean density is 500 and 2500 times the critical density of the Universe, respectively. The brightest cluster galaxy (BCG) location was determined by \citet{2023ApJ...947...44C} and is indicated with a red `X' in the right panel.


Our procedure for reducing the Chandra data and filtering out high-background periods follows the recommendations in the Chandra Analysis Guide\footnote{\url{https://cxc.harvard.edu/ciao/guides/acis_data.html}} and is described by \citet{Mantz1502.06020}, though we use more recent versions of the Chandra analysis software ({\sc ciao} 4.16) and calibration files ({\sc caldb} 4.11.0).
We additionally apply a time-dependent correction to the ancillary response files used in spectral analysis, motivated by an analysis of Chandra observations of Abell~1795 over the course of the mission (see details in \citealt{Mantz2512.05405}).

We assess the X-ray morphology of the cluster using the SPA metrics defined by \citet{Mantz1502.06020}, which quantify the sharpness of the surface brightness peak (``peakiness'', $p$) and the symmetry and alignment ($s$ and $a$) of elliptical isophotes corresponding to radii $\sim0.5$--1\,$r_{2500}$.
We find $s=1.20\pm 0.10$, $p=-0.417\pm 0.008$ and $a=1.29\pm 0.11$, satisfying the SPA criteria for relaxation and consistent with previous results from the earlier ACIS-I data alone \citep{2023ApJ...947...44C}.
The center of the X-ray emission from this analysis, approximately defined as the median photon position after background subtraction and flat-fielding, agrees with the position of the BCG to within $1''$.
We adopt the BCG position, $\alpha = 22^{\mathrm{h}}15^{\mathrm{m}}03\overset{\mathrm{s}}{.}9306$, 
$\delta = -35^{\circ}37\overset{\prime}{.}17\overset{\prime\prime}{.}885$
(J2000; \citealt{2023ApJ...947...44C}), as the cluster center in subsequent analysis.

For our spectral analysis, we use {\sc XSPEC}\footnote{\raggedright\url{http://heasarc.gsfc.nasa.gov/docs/xanadu/xspec/}} (version 12.12.1c) to model thermal emission from the hot ICM and the local Galactic halo as a sum of bremsstrahlung continuum and line emission components using the {\sc apec} plasma model and {\sc AtomDB}\footnote{\url{http://www.atomdb.org/}} version 3.0.9. The {\sc XSPEC} {\sc phabs} model was used to model photoelectric absorption by Galactic gas, using cross sections from \citet{1992ApJ...400..699B}. The hydrogen column density was fixed to $N_{\rm H}=1.00\times10^{20}\,\mathrm{cm}^{-2}$ \citep{HI4PI1610.06175}. We forward model the X-ray foreground, X-ray background and particle-induced background following the methods described in \citet{Mantz2512.05405}.

Our strategy for spectrally modeling the cluster follows that of \citet{Mantz1402.6212} and \Msixteen{}.
The cluster is modeled as a series of concentric, isothermal, spherical shells, with inner and outer radii determined by a defined set of annuli that span from the cluster center to the radius where the cluster signal represents a $2\sigma$ excess over the background in the 0.6--2.0\,keV energy band.
The expected signal in each annulus is computed by projecting this 3D model onto the plane of the sky.
In practice, this procedure allows for the densities in each shell to be constrained individually, while temperatures and metallicities must be linked between groups of adjacent shells. We fit the data nominally in the 0.6--7.0\,keV band, with some adjustments (following \citealt{Mantz2512.05405}) to limit the impact of background modeling uncertainties on our temperature measurements. In particular,
we discard spectral information, keeping only the integrated 1--2\,keV surface brightness, at radii $>49''$; figures showing temperatures/metallicities or derived quantities are truncated at this radius.

We perform two analyses of the cluster: one using the non-parametric {\sc projct} model in {\sc XSPEC} in which the ICM density and temperature profiles are a priori independent, and one in which they are related by the assumptions of hydrostatic equilibrium and a Navarro-Frenk-White (NFW) mass profile \citep{Navarro9611107} utilizing a modified version of the {\sc XSPEC} {\sc nfwmass} model (see \citealt{Mantz1402.6212}; \Msixteen{}).
For the latter, we follow \citet{Mantz1402.6212} by excluding a central region of radius $9''$ ($\sim75$\,kpc) from the analysis to prevent possible violations of hydrostatic equilibrium, spherical symmetry or the NFW mass model in this highest signal-to-noise region from disproportionately influencing the results.
This exclusion is modestly larger than the typical 50\,kpc radius used by default by \citet{Mantz1402.6212}, and was arrived at by iteratively increasing the exclusion radius until the best-fitting values of the NFW model stabilized. To evaluate the likelihood of spectral models, we used the Cash statistic \citep{Cash1979ApJ...228..939}, and a Markov Chain Monte Carlo (MCMC) algorithm was used to explore and constrain the credible regions of the model parameter space.


\section{Results and Discussion} \label{sec:analysis}

\noindent
The results of this analysis are presented in four subsections. Section \ref{sec:thermo_results} discusses the thermodynamic properties of SPT\,J2215, placing constraints on the cluster temperature, density, cooling time, and entropy profiles, and comparing the results to the sample of 40 lower redshift but similarly hot/massive, dynamically relaxed, cool-core clusters from \Msixteen{}. Section~\ref{sec:scaling_relations} presents scaling relations for the global thermodynamic quantities of SPT\,J2215 in the context of the same \Msixteen{} sample. Section \ref{sec:metalicity} presents and discusses the metallicity profile of the cluster.
In Section~\ref{sec:cosmo_results}, we present results on the X-ray emitting gas mass fraction ($f_\mathrm{gas}$) and mass concentration parameter, under the assumptions of hydrostatic equilibrium and an assumed NFW form for the mass profile. While the results in sections \ref{sec:thermo_results}, \ref{sec:scaling_relations} and \ref{sec:metalicity} do not assume hydrostatic equilibrium, they adopt reference radii $r_{2500}$ and $r_{500}$ determined from the mass modeling in Section~\ref{sec:cosmo_results}.


\begin{deluxetable*}{cccccccccc}
\tablecaption{Global properties of SPT-CL J2215-3537 
\label{tab:cluster_properties}}
\tablehead{
\colhead{$r_{2500}$} & 
\colhead{$r_{500}$} &
\colhead{$r_{200}$} & 
\colhead{$M_{500}$} & \colhead{$ f_{\rm gas}$} & \colhead{$c$}& \colhead{$M_{\rm gas}$} & \colhead{$kT_\mathrm {ce}$} & \colhead{$L$ (0.1-2.4\,keV)} & \colhead{$L_\mathrm{ce}$ (0.1-2.4\,keV)} \\ 
\colhead{(Mpc)} & \colhead{(Mpc)} & 
\colhead{(Mpc)} &
\colhead{($10^{14}\,M_\odot$)} & \colhead{} & \colhead{} &  \colhead{($10^{13}\,M_\odot$)} & \colhead{(keV)} & \colhead{($10^{45}$ erg s$^{-1}$)} & \colhead{($10^{45}$ erg s$^{-1}$)}
}
\startdata
$0.350^{+0.015}_{-0.013}$ & $0.79\pm0.06$ &
$1.20^{+0.13}_{-0.12}$ &
$5.2^{+1.4}_{-1.2}$ & $0.15\pm0.03$ & $4.3^{+1.3}_{-1.5}$ & $8.3\pm1.0$ & $9.2^{+2.3}_{-1.0}$ & $3.51^{+0.06}_{-0.09}$ &
$1.21^{+0.12}_{-0.13}$ \\
\enddata
\tablecomments{Measurements made from the {\sc nfwmass} analysis include \( r_{2500} \), \( r_{500} \), \( M_{500} \), \( c \), and \( f_{\rm gas} \), which was measured in a spherical shell spanning $0.8 < r/r_{2500} <1.2$. Measurements from the {\sc projct} analysis (i.e.\ \( M_{\rm gas} \), \( kT_{\rm ce} \), \( L \) and \( L_{\rm ce} \)) are made within \( r_{500} \). The X-ray temperature (\( kT_{\rm ce} \)) and center-excised luminosity (\( L_{ce} \)) are extracted from the region excluding the cluster core, specifically from 0.15–1 \( r_{500} \).}
\label{table:cluster_properties}
\end{deluxetable*}

\subsection{Thermodynamic Profiles} \label{sec:thermo_results}

\begin{figure*}[ht!]
    \centering
    \includegraphics[width=0.44\textwidth]{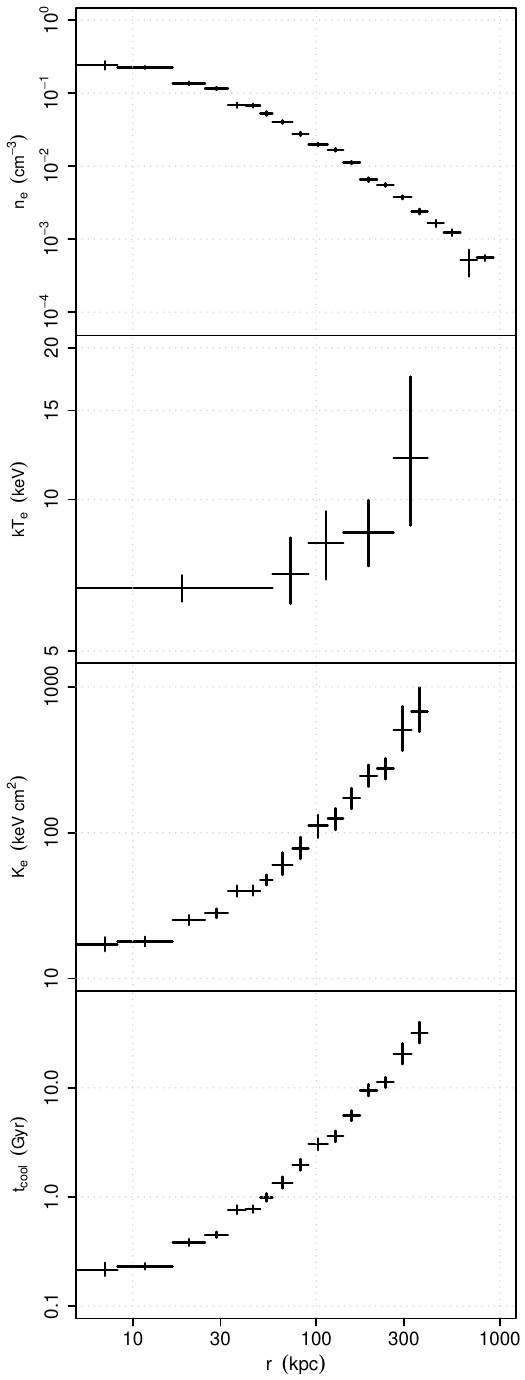}
    \includegraphics[width=0.44\textwidth]{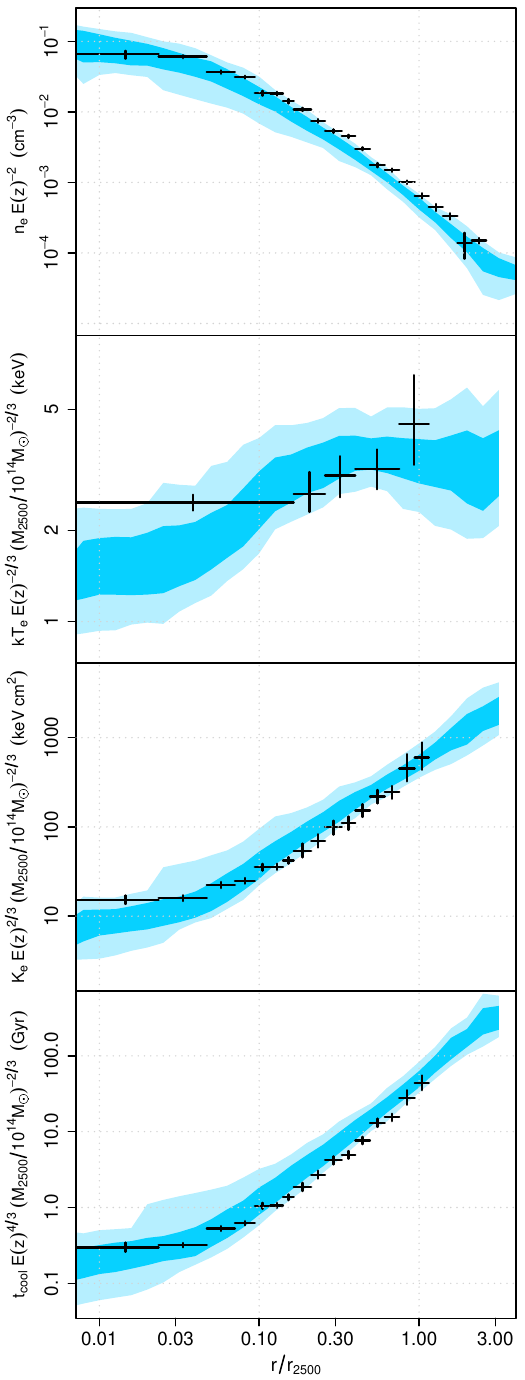}
    \caption{Thermodynamic profiles of SPT-CL J2215-3537 (black). From top to bottom are density, temperature, entropy, and cooling time. In the left panel, the profiles are presented without scaling. In the right panel, the 68.3\% and 95.4\% confidence ranges on profiles of a larger sample of relaxed, cool-core clusters over a redshift range of 0.078--1.063 (\Msixteen{}) are plotted in blue. The radii in the right panel have been scaled by $r_{2500}$, and the density, temperature, entropy, and cooling times have been scaled by $E(z)=H(z)/H_0$ and $M_{2500}$ as indicated by the corresponding axis label.
    \label{fig:thermo_profiles}}
\end{figure*}

\begin{figure}
    \centering
    \includegraphics[width=0.45\textwidth]{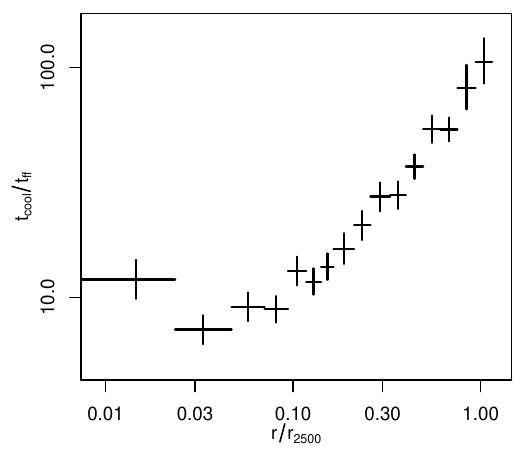}
    \caption{Radial profile of the cooling time, $t_{\rm cool}$, divided by the free-fall time, $t_{\rm ff}$, for SPT\,J2215. Indicative of the cool core with ongoing star formation, the ratio is consistent with a value of ten within $0.1\,r_{2500}$.  
    }
    \label{fig:tcooltff}
\end{figure}

\begin{figure}
    \centering
    \includegraphics[width=0.45\textwidth]{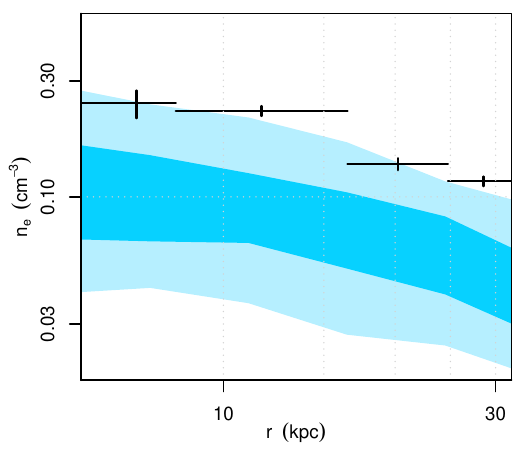}
    \caption{Unscaled central density profile of SPT\,J2215 (black) plotted against the 68.3\% and 95.4\% confidence ranges of the physical densities of the \Msixteen{} sample.  
    }
    \label{fig:physdens}
\end{figure}



\begin{figure*}[ht!]
\centering
\begin{minipage}[c]{0.9\textwidth} 
    \centering
    \includegraphics[width=0.48\textwidth]{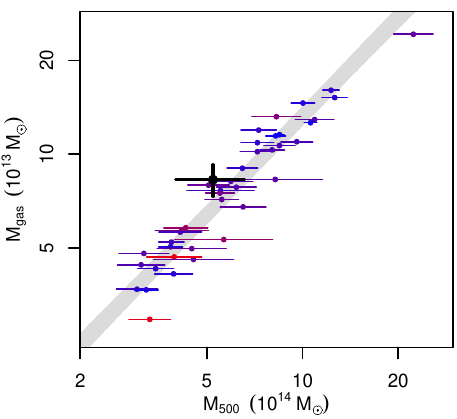}
    \includegraphics[width=0.48\textwidth]{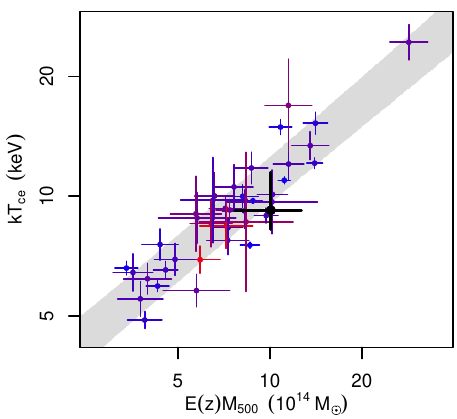}\\
    \includegraphics[width=0.48\textwidth]{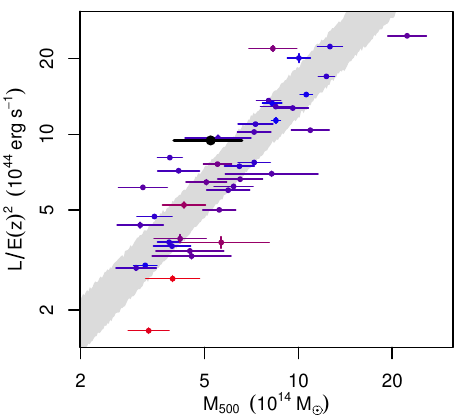}
    \includegraphics[width=0.48\textwidth]{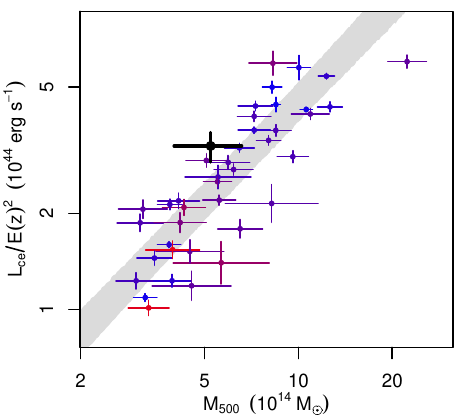}
\end{minipage}%
\hfill
\begin{minipage}[c]{0.08\textwidth} 
    \centering
    \includegraphics[width=\textwidth]{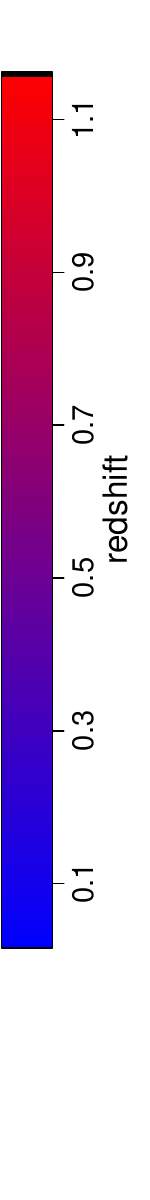}
\end{minipage}

\caption{Scatter plots of the integrated thermodynamic quantities for SPT\,J2215 (black) plotted against the rest of the relaxed cluster sample from \Msixteen{}. Shown are gas mass (top left), average temperature (top right), soft-band (0.1--2.4\,keV) luminosity (bottom left) and center-excised, soft-band luminosity (bottom right), all integrated within $r_{500}$. The average temperature and center-excised luminosity exclude radii $<0.15\,r_{500}$ (the former in 3D and the latter in projection). The \Msixteen{} clusters are color-coded by redshift according to the color bar. Shaded regions show the $1\sigma$ predictive range from a power-law fit to the \Msixteen{} data, including a log-normal intrinsic covariance.}
\label{fig:scaling_relations}
\end{figure*}

\noindent
Deprojected density and temperature profiles were obtained for SPT\,J2215 from the {\sc XSPEC} {\sc projct} analysis described in Section \ref{sec:methods}. The left column of Fig.\ \ref{fig:thermo_profiles} presents these unscaled profiles, as well as the derived profiles for pseudo-entropy and cooling time. In the right column, blue shading represents the 68.3\% and 95.4\% confidence ranges (encompassing both statistical uncertainty and intrinsic scatter) of the 40 cluster sample presented in \Msixteen{}, spanning a redshift range of 0.078--1.063. Radii in the right column are scaled by $r_{2500}$, and the profiles have been scaled to remove the redshift and mass dependence predicted by self similarity \citep{Kaiser1986MNRAS.222..323K}.
Though relatively extreme in some quantities, as discussed below, SPT\,J2215 is in general agreement with the lower-redshift, relaxed sample within the $2\sigma$ range.


The cool core of SPT\,J2215 can be clearly seen as its temperature decreases sharply towards the center from a maximum of approximately 12\,keV near $r_{2500}$ to a minimum value of approximately 6.7\,keV in the cluster center. 
Further evidence of the cool core can be seen in the entropy and cooling time profiles, pictured in the bottom two panels of Figure \ref{fig:thermo_profiles}, respectively. The pseudo-entropy is defined by the relation
\begin{equation} \label{eq:entropy}
K_e(r) = kT(r) n_e(r)^{-2/3},
\end{equation}
\citep{Balogh1999, Ponman9810359} where \(\ kT(r)\) is the measured radial temperature profile and \(\ n_e(r)\) is the measured electron density radial profile. 
The cooling time is similarly defined
\begin{equation} \label{eq:tcool}
t_\mathrm{cool}(r) = \frac{3}{2}\frac{(n_e(r) + n_p(r))kT(r)}{n_e(r) n_p(r) \Lambda(T(r),Z(r))}  
\end{equation}
where \( \Lambda(T(r),Z(r)) \) is the cooling function as defined by \citet{Sutherland1993} at temperature \( T \) and metallicity \( Z \), and \( n_e(r) \) and \( n_p(r) \) are the number densities of electrons and protons, respectively. Whether the central ICM is cooling rapidly enough to condense molecular clouds and form stars can be sensitively probed by threshold values of gas cooling time and entropy, shown to be $<1 \rm\,Gyr$ and $<30 \rm\,keV\,cm^2$, respectively \citep{Hudson0911.0409, 2012NJPh...14e5023M, Rafferty0802.1864, Cavagnolo2008ApJ...683L.107C, Pulido2018ApJ...853..177P, Calzadilla2311.00396}. The cooling time and entropy of SPT\,J2215 fall below these threshold values within $\sim60$\,kpc and $\sim30$\,kpc, respectively. 


Another quantity that is useful for characterizing the thermal properties of clusters is the free-fall time, \( t_{\rm ff} = (2r/g)^{1/2} \), where \( g \) is the local gravitational acceleration at radius \( r \). The free-fall time defines the characteristic timescale of gravitational collapse of gas in the ICM, and, correspondingly, the ratio \(t_{\rm cool}/t_{\rm ff}\) quantifies the thermal stability of the gas against gravity. When this ratio reaches levels of $\sim10$--20, thermal instability and subsequent cooling may lead to the formation of cold, molecular gas and star formation \citep{Gaspari2012, Sharma2012, Voit1409.1598, Hogan1704.00011, Pulido2018ApJ...853..177P}. We see in Figure\ \ref{fig:tcooltff} that the ratio for SPT\,J2215 is consistent with values of $\sim10$ within $\sim 0.1\,r_{2500}$ ($\sim 35$\,kpc). This suggests that the core of SPT\,J2215 is thermally unstable, potentially leading to enhanced central gas density and ongoing star formation. Indeed, while consistent at the $\sim 2\sigma$ level, the central physical (unscaled) gas density for SPT\,J2215 lies at the upper edge of the distribution, relative to the 40 cluster sample of \Msixteen{} (shown in Fig.\ \ref{fig:physdens}), hinting at the presence of excess X-ray emission from cooling gas. Interestingly, at $320^{+230}_{-140}\,\rm M_\odot\,yr^{-1}$ \citep{2023ApJ...947...44C}, the star formation rate at the center of SPT\,J2215 is also more rapid than many of the strongest known cooling clusters \citep{McDonald1803.04972, Calzadilla_2022}. 

Another mechanism which may give rise to enhancements in the inferred central gas density in the presence of ongoing star formation is the turbulent mixing of hot and cold gas phases which, like cooling, can lead to the presence of X-ray bright multiphase gas. Here, in principle, the mixing motions could be sourced by the central AGN. 

While the peaked X-ray surface brightness profiles of cool-core clusters are typically associated with enhanced central gas densities, it has recently been speculated that inverse Compton (IC) scattering of cosmic microwave background (CMB) photons off cosmic ray electrons associated with central radio sources may also contribute to the central X-ray flux \citep{Hopkins2025}. 
Given the relatively high redshift of SPT\,J2215 and the expected $(1+z)^4$ scaling of the emissivity for this process, some contribution to the central X-ray emission from SPT\,J2215 may be possible, but testing this conjecture would require deep radio observations. However, the high-quality data available for low redshift, cool-core systems such as the Perseus Cluster--which hosts both an X-ray bright cool core and radio bright central radio mini halo (\citealt{vanWeeren2024} and references therein) but shows neither spatially extended non-thermal X-ray emission (see for example the emission line ratios presented by \citealt{HitomiPerseusAbund2017}) nor detectable, diffuse very high energy gamma ray flux (\citealt{Ahnen2016})--suggest that this emission process is likely more important for exceptionally powerful radio sources in lower-mass groups at even higher redshifts (e.g.\ \citealt{Fabian0301468, Erlund2006MNRAS.371...29E}).

\subsection{Scaling Relations}
\label{sec:scaling_relations}

\noindent
Scaling relations of integrated cluster properties provide a more global basis for comparing SPT\,J2215 with the \Msixteen{} sample.
Figure~\ref{fig:scaling_relations} shows this comparison, specifically for gas mass, average temperature, soft-band (0.1--2.4\,keV) luminosity and center-excised, soft-band luminosity integrated within $r_{500}$ (our measurements for SPT\,J2215 are summarized in Table~\ref{tab:cluster_properties}).
For consistency with \Msixteen{}, the average temperature and center-excised luminosity both exclude radii $<0.15\,r_{500}$ (the former in 3D and the latter in projection).
In the figure, points are color-coded according to the colorbar by increasing redshift from blue to red, except for SPT\,J2215 which is shown in black, while gray shading shows predictive ranges from a joint power-law fit from \Msixteen{}, including a log-normal intrinsic covariance of the four quantities.
Analogously to the results discussed above, we see that SPT\,J2215 is generally consistent with the lower-redshift relaxed clusters, while also being relatively high in gas mass and luminosity, both following from its high gas density.
We note that this does not also apply for the other $z>1$ clusters shown (red points), suggesting that our measurements for SPT\,J2215 are more the result of its own formation history than a broader evolutionary trend within the relaxed sample.

\subsection{Metal Enrichment} 
\label{sec:metalicity}

\begin{figure}
    \centering
    \includegraphics[width=0.45\textwidth]{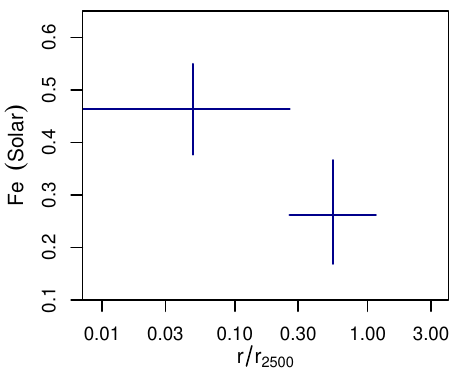}
    \caption{Iron abundance (expressed as a fraction of solar abundance) profile for SPT\,J2215 scaled by $r_{2500}$, showing an enhanced central metallicity which decreases outside of the core to be consistent with the universal value of $\sim0.3$\,\(\textup{Z}_\odot\). 
    }
    \label{fig:metal}
\end{figure}
\noindent
Figure \ref{fig:metal} presents the deprojected ICM iron abundance profile of SPT\,J2215 in solar units \citep{Asplund0909.0948}, where abundances were measured from the equivalent width of the Fe K$\alpha$ emission line. The profile indicates a gradient in metallicity, which decreases from a moderately peaked central value of $0.46\pm0.09$\,\(\textup{Z}_\odot\) to $0.26\pm0.09$\,\(\textup{Z}_\odot\) outside of the core. 
An enhanced central metallicity has been observed in samples of relaxed clusters at redshifts $z\lesssim1$ \citep{1998MNRAS.297L..63A, 2004A&A...419....7D, 10.1093/mnras/stx2200}, as well as multiple relaxed clusters at $z\gtrsim1$ (e.g.\ \citealt{Santos1111.3642, 10.1093/mnras/stx2200, Liu2018MNRAS.481..361L}). The measured outer abundance of SPT\,J2215 is consistent with the ``universal" level of $\sim0.3$\,\(\textup{Z}_\odot\) observed at intermediate-to-large radii in lower redshift systems \citep{Ettori1504.02107, McDonald1603.03035, Liu200312426}. As a relaxed cluster at redshift $z>1$, both the presence of a gradient in its metallicity profile and the suggestion that its abundance settles to $\sim0.3$\,\(\textup{Z}_\odot\) outside of the core would imply that the enrichment of the gas observed in the outskirts of SPT\,J2215 and systems at lower redshifts occurred at early times, in the protocluster phase (see also \citealt{Flores2108.12051}).   

\subsection{Cosmology} 
\label{sec:cosmo_results}

\begin{figure*}
    \centering
        \centering
        \includegraphics[scale=1.2]{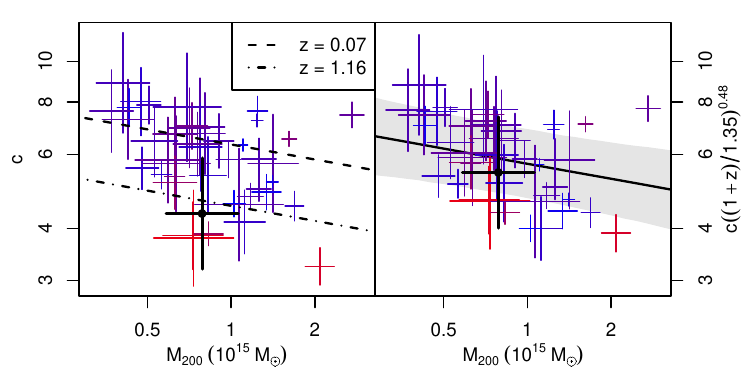}
        \centering
        \includegraphics[scale=1.2]{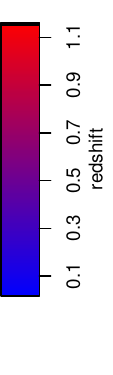}
    \caption{Concentration--mass measurements for members of the relaxed cluster sample \citep{Darragh-Ford2302.10931}. SPT\,J2215 is plotted in black, while the rest of the sample is color-coded from blue to red in order of increasing redshift. {\it Left:} The best-fitting power law model of concentration as a function of mass and $1+z$ from \citet{Darragh-Ford2302.10931}, evaluated at $z=0.07$ and $z=1.16$, is shown using dashed and dot-dash lines. {\it Right:} Concentration--mass relation with the best-fitting dependence on $1+z$ divided out. The gray shaded region indicates the 68.3\% posterior predictive interval, accounting for intrinsic scatter.
    }
    \label{fig:concentrationmass}
\end{figure*}

\begin{figure}
    \centering
        \centering
        \includegraphics[scale=1.0]{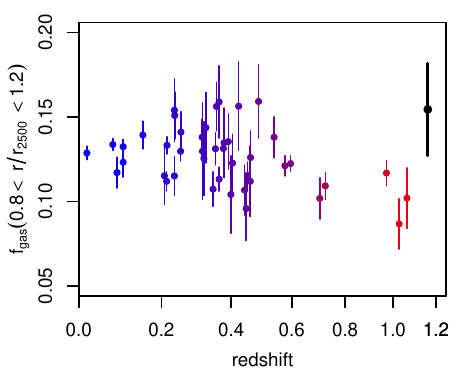}
    \caption{Gas mass fraction ($f_{\rm gas}$) plotted as a function of redshift for members of the relaxed cluster sample in \Mtwentytwo{} and \DFtwentythree{}. New constraints obtained from this work on SPT\,J2215 are designated with a filled black point, while the rest of the sample is color-coded in redshift according to the same colorbar as in Figure \ref{fig:concentrationmass}.
    }
    \label{fig:fgas}
\end{figure}

\noindent
In the following analysis, we use the {\sc nfwmass} model in {\sc XSPEC}, with the assumptions of an NFW mass profile and hydrostatic equilibrium. This allows us to place constraints on the gas mass fraction, $f_{\rm gas}$, (below) and the NFW concentration parameter, $c=r_{200}/r_s$. The left panel of Figure \ref{fig:concentrationmass} plots the constraints obtained in this analysis for $M_{200}$ and the concentration of SPT\,J2215 (black) against the constraints obtained for the 44 cluster sample from \DFtwentythree{}, which are colored from blue to red by increasing redshift. Shown as dashed and dot-dash lines are best fit power law model predictions in mass and $1+z$ from \DFtwentythree{} for redshifts of $z=0.07$ and $z=1.16$, respectively. The right panel of Figure \ref{fig:concentrationmass} plots the concentrations with the best-fitting $1+z$ power law from \DFtwentythree{} divided out. The now redshift-independent best-fitting concentration--mass relation is plotted as a solid black line, with its 68\% posterior predictive interval, including intrinsic scatter, indicated by the gray shaded region. Our measurement of the concentration of SPT\,J2215, $c=4.3^{+1.3}_{-1.5}$, is consistent with the model prediction from the lower redshift data.


As is common in the literature, we measure a gas mass fraction, $f_{\rm gas}=M_{\rm gas}/M_{\rm tot}$, integrated within the characteristic radius $r_{500}$ to be $f_{\rm gas}=0.15\pm0.03$. We find a nearly identical value in the spherical shell spanning $0.8$--$1.2\,r_{2500}$, $f_{\rm gas}=0.15\pm0.03$. We report this value in Table~\ref{tab:cluster_properties}. This gas mass fraction is plotted with 68.3\% confidence bounds in Figure \ref{fig:fgas} with the $f_{\rm gas}$ measurements of the relaxed cluster sample of \Mtwentytwo{}, which follow the same color scaling in redshift as in Figure \ref{fig:concentrationmass}. 
While our constraints for SPT\,J2215 are not substantially more precise than those of \Mtwentytwo{}, due to our more careful exclusion of the cluster center when assuming hydrostatic equilibrium, they remain consistent with the expectation of a constant gas mass fraction in the concordance model.

\section{Conclusion} \label{sec:conclusion}

\noindent
With 132\,ks of new \Chandra{} observations in addition to 67\,ks of archival data, we provide updated constraints on the thermodynamic, metal enrichment, and cosmological properties of the $z=1.16$ relaxed, cool-core cluster SPT-CL J2215-3537. Our conclusions are summarized as follows:

\begin{enumerate}  
    \item Though an extreme example in terms of its redshift, the measured temperature, entropy, density, and cooling time profiles of SPT\,J2215, scaled according to self-similar models, are consistent with those of lower-redshift, cool-core clusters. We do note, however, that at the $\sim 2\sigma$ level the central physical gas density in SPT\,J2215 may be higher than that of its lower-redshift counterparts.
    \item The metallicity profile of SPT\,J2215 indicates the presence of a gradient with a peaked central metallicity, a feature that has also been observed in at least one other $z > 1$ relaxed cluster (\citealt{Santos1111.3642, Liu2018MNRAS.481..361L}). The abundance of SPT\,J2215 approaches $\sim0.3$\,\(\textup{Z}_\odot\) outside of the cluster core, consistent with the universal level of enrichment at large radii found in studies at lower redshifts.
    \item We measure an NFW concentration parameter of $c=4.3^{+1.3}_{-1.5}$ and a gas mass fraction of $f_{\rm gas}=0.15\pm0.03$ within a spherical shell spanning $0.8$--$1.2\,r_{2500}$, identical to the value of $f_{\rm gas}$ measured within the
    characteristic radius $r_{500}$. These measurements are fully consistent with $\Lambda \text{CDM}$ model predictions.    
\end{enumerate}

Our work extends previous studies of the thermodynamics and cosmology of relaxed galaxy clusters to higher redshifts, supporting the conclusion that such systems can be well described by simple hydrodynamic models, with regulated feedback in their centers, and with matter distributions consistent with $\Lambda \text{CDM}$ model predictions.
While very few massive, relaxed clusters are currently known at $z\gtrsim1$, the discovery of SPT\,J2215 in SPT-ECS and, more recently, the $z=1.5$ relaxed cluster ACT-CL\,J0123.5-0428 in the Advanced ACTPol survey (Flores et al., in preparation), demonstrate the potential to expand such studies by following up Sunyaev--Zel'dovich effect detections.
At these redshifts, Chandra and XMM-Newton are still able to identify and characterize sources of interest, albeit less efficiently than at lower redshifts.
In the longer term, new X-ray facilities such as NewAthena or AXIS should be able to routinely make such measurements out to $z\sim2$ and beyond.

\section*{Acknowledgments}
    This research has made use of data obtained from the Chandra Data Archive provided by the Chandra X-ray Center (CXC). Support for this work was provided by the National Aeronautics and Space Administration through Chandra Award Number GO2-23113A issued by the Chandra X-ray Center, which is operated by the Smithsonian Astrophysical Observatory for and on behalf of the National Aeronautics Space Administration under contract NAS8-03060. We acknowledge support from the U.S. Department of Energy under contract number DE-AC02-76SF00515, and support from NASA Astrophysics Data Analysis award 80NSSC24K1401.

    This work was performed in the context of the South Pole Telescope scientific program. The South Pole Telescope program is supported by the National Science Foundation (NSF) through awards OPP-1852617 and OPP-2332483. Partial support is also provided by the NSF Physics Frontier Center grant PHY-0114422 to the Kavli Institute of Cosmological Physics at the University of Chicago, the Kavli Foundation and the Gordon and Betty Moore Foundation grant GBMF 947 to the University of Chicago. The SPT is also supported by the U.S. Department of Energy. Argonne National Laboratory’s work was supported by the U.S. Department of Energy, Office of High Energy Physics, under contract DE-AC02-06CH11357.

    Following acceptance of this work for publication, an independent study of the same Chandra data was posted on arXiv by \citet{Bessa2026ApJ}.
  

\facility{CXO}


\software{
  Astropy \citep{astropy1304.002, Astropy-Collaboration1307.6212, Astropy-Collaboration1801.02634, Astropy-Collaboration2206.14220},
  CIAO \citep{ciao1311.006, Fruscione2006SPIE.6270E..1VF},
  HEASOFT \citep{heasoft1408.004},
  LMC \citep{lmc1706.005},
  MARX \citep{marx1302.001, Davis2012SPIE.8443E..1AD},
  SXRBG \citep{sxrbg1904.001},
  XSPEC \citep{xspec9910.005, Arnaud1996ASPC..101...17A}
}


\bibliographystyle{aasjournal}
\newcommand{\asl}{ASL} 
\bibliography{references,newrefs,main} 

\end{document}